\documentclass[useAMS,usenatbib]{mn2e}
\usepackage{graphicx,amssym}
\newcommand{\bc}{\begin{center}}
\newcommand{\ec}{\end{center}}
\newcommand{\bow}{{\small BOW06 }}
\newcommand{\dlb}{{\small DLB07 }}
\newcommand{\guo}{{\small GUO11 }}

\title[On the stellar populations of massive galaxies]
      {On the stellar populations of massive galaxies}  
\author[G.~De Lucia \& S. Borgani]
       {Gabriella De Lucia$^{1}$\thanks{Email: delucia@oats.inaf.it}, 
        and Stefano Borgani$^{2,1,3}$\\
        $^1$INAF - Astronomical Observatory of Trieste, via G.B. Tiepolo 11, 
        I-34143 Trieste, Italy\\
       $^2$Dipartimento di Fisica, Universit\`a degli Studi di Trieste, via
       A. Valerio 2, Trieste, Italy\\ 
       $^3$INFN, Istituto Nazionale di Fisica Nuclare, via Valerio 2, I-34127
       Trieste, Italy} 
\begin{document}

\pagerange{\pageref{firstpage}--\pageref{lastpage}} 
\pubyear{2012}

\maketitle

\label{firstpage}

\begin{abstract}
In this Letter, we analyse the predicted physical properties of massive
galaxies, in the framework of recent semi-analytic models of galaxy formation.
All models considered account for winds driven by supernovae explosions and
suppression of gas condensation at the centre of relatively massive haloes by
active galactic nuclei (AGN). We show that, while these models successfully
reproduce the old stellar populations observed for massive galaxies, they fail
in reproducing their observed chemical abundances. This problem is alleviated
but still present if AGN feedback is completely switched off. Moreover, in this
case, model predictions fail in accounting for the old stellar ages of massive
galaxies. We argue that the difficulty of semi--analytical models in
simultaneously reproducing the observed ages and metallicities of massive
galaxies, signals a fundamental problem with the schemes that are currently
adopted to model star formation, feedback, and related recycling of gas and
metals.
\end{abstract}


\section{Introduction}
\label{sec:intro}

In the current standard picture for structure formation, galaxies originate
from gas condensation within the potential wells of hierarchically formed dark
matter haloes. The mass distribution of these haloes is well described by a
power-law, differing significantly from the typical shape of the observed
galaxy mass function.

Stellar feedback is believed to play a crucial role at the low mass regime
\citep{White_and_Frenk_1991, Benson_etal_2003}. Observations suggest that
galactic-scale outflows are ubiquitous in starburst galaxies at all cosmic
epochs, and that the outflowing material is multiphase (containing cold, warm,
and hot gas, dust and magnetized relativistic plasma -
\citealt[e.g.][]{Heckman_2002,Weiner_etal_2009}). Unfortunately, available
observational measurements refer to material that is still relatively deep
within the gravitational potential of the halo. So the estimated outflow rates
are difficult to translate into rates at which mass, metals, and energy escape
from galaxies and are eventually transported into the inter-galactic
medium. The fate of the outflowing material will depend critically on a number
of unknowns, as well as on its multiphase nature. Given the uncertainties, it
is not clear how appropriate the different prescriptions adopted for treating
galactic winds in galaxy formation models are.

At the massive end, feedback from active galactic nuclei (AGN) is believed to
play a key role. X--ray observations show that AGN feedback should be
responsible for the thermal structure of ``cool cores''; only a modest amount
of gas cools and form stars at the centre of galaxy clusters, despite the short
cooling timescales inferred from the X-ray emission.  Detailed recent studies
have confirmed that brightest cluster galaxies (BCGs) are more likely to host a
radio-loud AGN than other galaxies of similar stellar mass, and have shown that
the ensemble averaged power from radio galaxies is sufficient to offset the
mean level of cooling
\citep[e.g.,][]{Best_etal_2007,mcnamara_nulsen07}. Heating from a central AGN
has become a crucial ingredient for galaxy formation models in order to
reproduce the observed exponential cut-off at the bright end of the galaxy
luminosity function, and the old stellar populations observed for massive
galaxies \citep[e.g.][]{Croton_etal_2006,DeLucia_etal_2006}. The details of the
energy injection by the central engine, and coupling with the surrounding hot
gas are, however, still unclear. So the prescriptions adopted to model this
process are very schematic, and often not well grounded in observations.

\section{The galaxy formation models}
\label{sec:simsam}

In this study, we take advantage of publicly available galaxy
catalogues\footnote{These are available at the following webpage:
  http://www.mpa-garching.mpg.de/millennium/}, constructed applying
semi-analytic methods to high-resolution cosmological simulations. In
particular, we use catalogues from the models by \citet[][\bow
  hereafter]{Bower_etal_2006}, \citet[][\dlb]{DeLucia_and_Blaizot_2007}, and
\citet[][\guo]{Guo_etal_2011} applied to the Millennium Simulation
\citep{Springel_etal_2005}.

The first two models have been developed independently and differ in a number
of details related to the merger tree construction and the prescriptions
adopted for various physical processes considered. The \guo model is based on
the \dlb one, but it adopts a different (significantly more efficient)
supernovae (SN) feedback and an improved treatment for the evolution of
satellite galaxies. All three models include prescriptions for SN driven winds,
follow the growth of super-massive black holes, and include a phenomenological
description of AGN feedback. All three models adopt an instantaneous recycling
approximation, and assume an efficient mixing of metals within the cold gaseous
phase.

Using the public catalogues, we have selected all galaxies in a sub-cube of the
Millennium Simulation of $100\, {\rm Mpc}$ on a side. In addition, we have
selected all central galaxies of haloes more massive than\footnote{$M_{200}$ is
  the mass contained within the radius encompassing an average density of 200
  times the critical cosmic density.}  $M_{200}=5\times 10^{14}\,{\rm
  M}_{\sun}$. We focus, in particular, on the stellar ages and metallicities of
model galaxies. Unfortunately, the public catalogues do not provide consistent
information: ages and metallicities are weighted by mass in the \dlb and \guo
models, and by V-band luminosity for the \bow model. This does not affect,
however, our conclusions. In the following, we will assume for the solar
metallicity: $Z_{\odot} = 0.02$.

\section{The stellar populations of model galaxies}
\label{sec:sp}

\begin{figure*}
\bc
\resizebox{18cm}{!}{\includegraphics{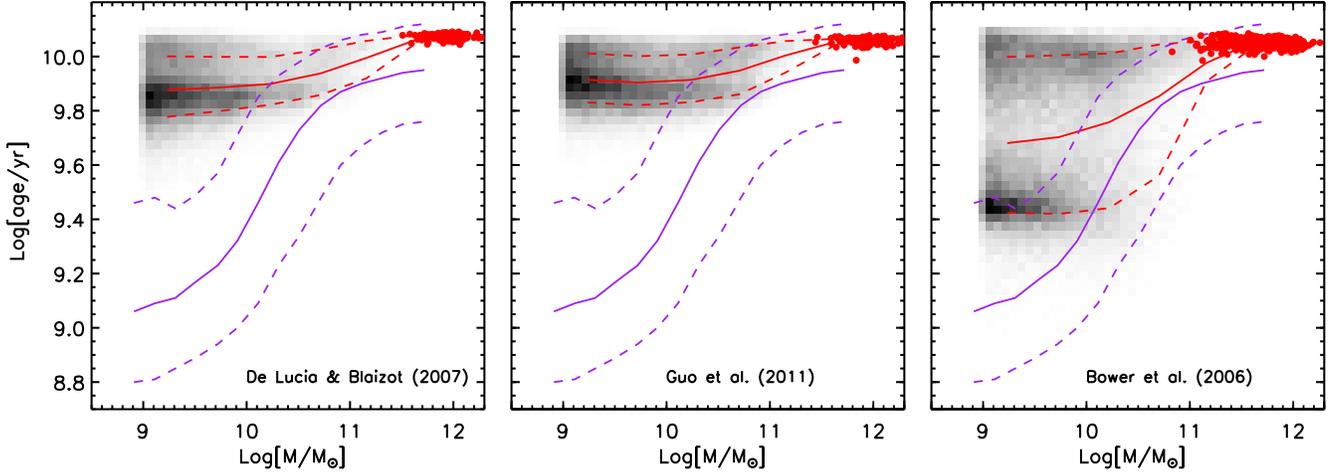}}\\%
\caption{Age-stellar mass relation for the three models considered. Grey maps
  show the distribution of model galaxies in a sub-volume of the Millennium
  Simulation of size $100\,{\rm Mpc}$ on a side. Red solid lines show the
  median, while dashed red lines mark the 16th and 84th percentiles of the
  distributions. Filled circles show the location of the central galaxies of
  haloes with $M_{200}>5\times 10^{14}\,{\rm M}_{\sun}$.  The \dlb and the \guo
  model ages are weighted by mass, while the \bow model ages are weighted by
  V-band luminosity. Purple lines show the conditional distribution of stellar
  ages as a function of stellar mass from \citet{Gallazzi_etal_2005}.}
\label{fig:age_mass}
\ec
\end{figure*}

\begin{figure*}
\bc
\resizebox{18cm}{!}{\includegraphics{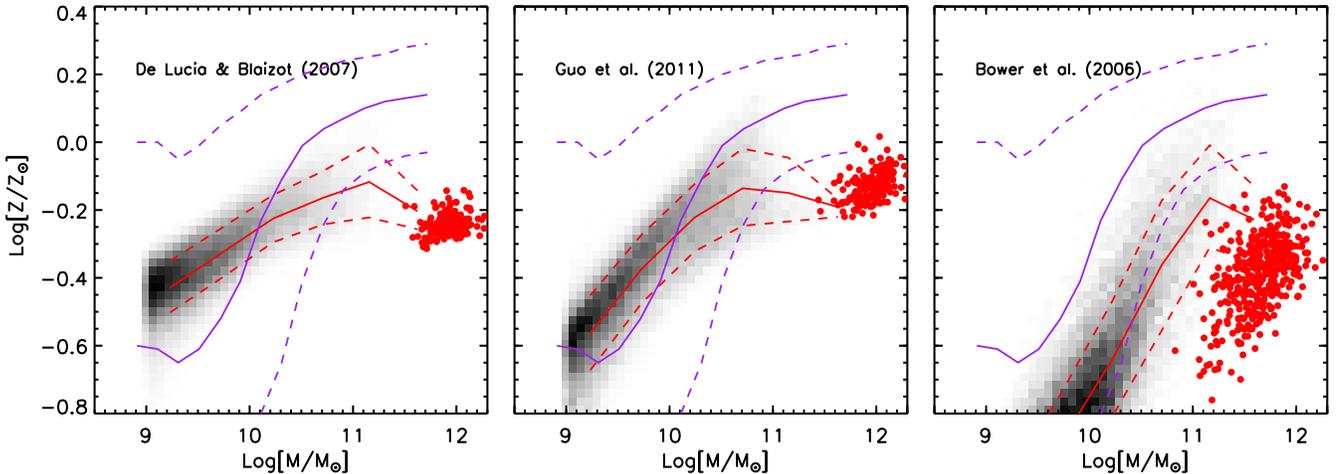}}\\%
\caption{As in Fig.~\ref{fig:age_mass} but for the metallicity-stellar mass
  relation. The \dlb and the \guo metallicities are weighted by mass, while the
  \bow metallicities are weighted by V-band luminosity. Lines and symbols have
  the same meaning as in Figure~\ref{fig:age_mass}.}
\label{fig:met_mass}
\ec
\end{figure*}

The gray regions in Figure \ref{fig:age_mass} show the age-stellar mass
relation for all model galaxies in the sub-volume of the Millennium simulation
considered. Solid and dashed lines show the median and the 16th and 84th
percentiles of the distributions, while filled symbols show the location of
central galaxies of haloes with $M_{200}>5\times 10^{14}\,{\rm M}_{\sun}$. As
mentioned above, the stellar ages from the \dlb and \guo models are weighted by
mass, while those from the \bow model are weighted by V-band luminosity. The
latter are more directly comparable to the observational measurements by
\citet{Gallazzi_etal_2005}, shown as purple lines. These are obtained adopting
a Bayesian statistical approach to derive full likelihood distributions for
ages and metallicities by comparing the observed spectra to a library of model
star formation histories. The comparison is based on the strengths of spectral
absorption features that depend weakly on the [$\alpha$/Fe] ratio. The
distributions obtained for model galaxies are not convolved by observational
uncertainties, that are relatively large particularly for older ages.

Model galaxies tend to be on average older than observational estimates,
particularly for low-mass galaxies. This problem is in part due to the well
known excess of faint and passive model satellites \citep[e.g.][and references
  therein]{Wang_etal_2007,Weinmann_etal_2010}. It appears less severe when
considering luminosity-weighted ages (see right panel), but models still miss a
population of very young low-mass galaxies. At the massive end, model galaxies
are dominated by very old stellar populations so that differences between
mass-weighted and luminosity-weighted ages are small. In this mass range, model
predictions are close to the upper limits of the observational
estimates. Central galaxies of massive clusters have very old ages: in the \dlb
model, the median age is $11.8$~Gyr with a scatter of only about $0.3$~Gyr. For
the same galaxies, the \guo model predicts somewhat younger mass-weighted ages
(the median age is $11.4$~Gyr) with a comparable scatter. This is likely due to
the fact that, in this model, the hot reservoir associated with satellite
galaxies is not stripped instantaneously. Cooling is allowed on these galaxies
so that they are on average more gas-rich than in a model with instantaneous
stripping of hot gas. Their accretion onto the central galaxies trigger star
formation episodes that slightly rejuvenate the stellar population of the
remnant galaxies. For these galaxies, the median luminosity weighted age from
the \bow model is $11.2$~Gyr, with a $\sim 0.4$~Gyr scatter.

Figure~\ref{fig:met_mass} shows model predictions and observational estimates
for the stellar metallicity-mass relation, with lines and symbols having the
same meaning as in Figure~\ref{fig:age_mass}. All models predict a relatively
tight metallicity-mass relation with a steep slope and a pronounced turn-over
at the most massive end. Central galaxies of massive clusters have stellar
metallicity of $\sim 0.57\,Z_\odot$ in the \dlb model, with a $\sim 0.04$
scatter. Because of the more efficient supernovae feedback, the \guo model
predicts a steeper metallicity-mass relation. For the central massive galaxies
considered, this model predicts a median metallicity of $\sim 0.74\,Z_\odot$
with a $\sim 0.08$ scatter. The luminosity weighted metallicities from the \bow
model are offset low with respect to predictions from other models, as
expected. For the central galaxies of massive clusters, this model predicts a
median metallicity of $\sim 0.43\,Z_\odot$, with a $\sim 0.10$ scatter. Purple
lines show measurements by \citet{Gallazzi_etal_2005}. The typical uncertainty
on stellar metallicity is $\sim 0.12$~dex, but there is a significant tail of
galaxies with uncertainties of up to $\sim 0.25$~dex. Low-metallicity galaxies
tend to be associated with larger uncertainties because of the weaker
absorption lines.

\section{The stellar populations of BCGs}
\label{sec:bcgs}

The models used in this study successfully reproduce the old stellar
populations observed for massive galaxies. If any, model galaxies are {\it too
  old} with respect to data. All three models considered, however, fail in
reproducing the observed chemical abundances of massive galaxies: at the
massive end, the predicted metallicity mass relation turns-over and is offset
low with respect to the data. The problem appears to be more severe when
considering the most massive galaxies at the centre of relatively massive
haloes, but is not limited to them. This is due to the fact that many of the
most massive satellites have been accreted recently from relatively massive
haloes \citep{DeLucia_etal_2012}. Therefore, the same physical processes acting
on today's BCGs have played an important role during the lifetime of these
galaxies.

\begin{figure*}
\bc
\resizebox{18cm}{!}{\includegraphics{figures/check_noagn}}\\%
\caption{Distribution of stellar masses (left), mass-weighted ages (middle),
  and stellar metallicities (right) for central galaxies of haloes with mass
  $M_{200}>5\times 10^{14}\,{\rm M}_{\sun}$. Black unfilled histograms
  correspond to the fiducial model described in \dlb, while red hatched ones
  correspond to the same model but with AGN feedback switched off. Purple cross
  hatched histograms correspond to a model with AGN feedback but the SN
  feedback scheme used in \citet{DeLucia_etal_2004}.}
\label{fig:check_noagn}
\ec
\end{figure*}

From the observational viewpoint, while stellar populations in early type
galaxies have been extensively studied, very little is known about the stellar
populations of BCGs. In a recent work, \citet{vonderLinden_etal_2007} studied a
sample of $\sim 600$ BCGs and contrasted their stellar populations to those of
a `control-sample' of non-BCGs matched in stellar mass, redshift and
colour. They found that the ages and the metallicities of BCGs are comparable
to those typical of massive ellipticals, while the $\alpha$-enhancements are
higher in BCGs. \citet{Loubser_etal_2009} analysed the stellar populations of
$\sim 50$ nearby BCGs using long-slit spectroscopy. They found that most BCGs
are very old, but some (generally associated with cooling flows clusters) show
signatures of recent star formation. The metallicity and $\alpha$-enhancement
distributions measured by Loubser et al. peak at values {\it higher} than those
obtained for normal giant ellipticals. It is worth noting that these
measurements apply to the central regions of BCGs: the inner 3 arcsec in the
study by von der Linden et al., and 1/8 of the half-light radius in Loubser et
al. However, the mean stellar population gradients of BCGs (null for
$\alpha$-enhancements, almost null for age, and negative\footnote{Around -0.3
  dex per decade of variation in the radius.} for metallicity) are consistent
with those of normal massive elliptical galaxies
\citep{Loubser_and_Sanchez-Blazquez_2012}. Therefore, any attempt to correct
for aperture effects would change the normalization but not the shape of the
model mass-metallicity relation at the massive end, making the discrepancy
shown in Fig.~\ref{fig:met_mass} robust and severe.

In models where cooling flows are suppressed at late times by AGN activity
(like those considered here), the stars of massive central galaxies form early,
in low-mass progenitors whose high star formation rates are fuelled by rapid
cooling. Galaxies are then assembled relatively late, through the accretion of
many smaller satellites, driven by the merging history of the parent halo. The
metallicity-mass relation does not evolve significantly in the models, so that
these late dry mergers tend to {\it decrease} their total stellar
metallicity. Dry minor mergers would likely deposit metal poor stars in the
outer regions of the remnant, so they would not affect significantly the
central stellar metallicity of BCGs. For the \dlb model, we find that the
stellar metallicity of the main progenitor of the BCGs at $z\sim 1$ is on
average only about 3 per cent lower than its final value, so that even removing
all stars accreted after $z\sim 1$ does not improve the disagreement with
observational data.  Therefore, in a scenario where the mass growth of the most
massive galaxies is dominated by minor dry mergers, it appears difficult to
reproduce simultaneously the observed mass-metallicity relation and the
chemical abundances of the most massive galaxies.

It is interesting to analyse what happens when AGN feedback is switched off. To
this aim, we have re-run the \dlb model on all merger trees of haloes more
massive than $5\times 10^{14}\,{\rm M}_{\sun}$. Fig.~\ref{fig:check_noagn}
shows the distribution of stellar masses, mass-weighted ages, and stellar
metallicities for all central galaxies of these haloes. Black unfilled
histograms show results from the \dlb model, while red hatched ones correspond
to the same model but with no AGN feedback. As expected, galaxies become
significantly more massive (by a factor $\sim 6$) in a model with no AGN
feedback. Gas cooling and, in minor part, accretion of other gas-rich
satellites provide fresh material for late star formation, which makes the
stellar population of these galaxies significantly younger (by $\sim
5$~Gyr). Since the material out of which new stars form is enriched by past
generations of stars, the mean stellar metallicity increases going from $\sim
0.6\,Z_\odot$ to $\sim 0.7\,Z_\odot$. The increase is not large because the
metallicity of the cooling gas is relatively low ($\sim 0.1-0.2\,Z_\odot$
solar, in agreement with observational measurements), and it is further diluted
by metals being deposited through accreted galaxies.  Interestingly, a recent
study by \citet{McCarthy_etal_2010} shows that hydrodynamical simulations
including AGN feedback provide central group galaxies that have too low
metallicity with respect to observational data. They find, however, that when
AGN feedback is switched off, the metal content of central galaxies increases
dramatically, becoming even larger than observed.

\section{Discussion and Conclusions}
\label{sec:discconcl}

All recent models of galaxy formation combine a relatively strong SN feedback
with a model for AGN feedback to suppress cooling at the centre of relatively
massive haloes.  The former is needed to bring the faint-end slope of the
galaxy luminosity/mass function in agreement with the relatively shallow value
measured. The latter affects significantly the number densities and stellar
ages of the most massive galaxies. Our observational and physical understanding
of both processes is rather poor so that both are described in galaxy formation
models using quite schematic prescriptions.

We have shown that models fail dramatically in reproducing the observed stellar
populations of galaxies. At the low mass end, they tend to over-predict the
fraction of passive galaxies, and lack a population of very young galaxies that
is observed. This is a well known problem, that plagues {\it all} recently
published models as well as hydrodynamical simulations
\citep{Weinmann_etal_2012}. In this letter, we have shown that significant
discrepancies are found also for the most massive galaxies, whose model stellar
populations are very old but relatively metal poor.

As mentioned above, the modelling adopted for AGN feedback is quite simple and
neglects some relevant aspects of AGN activity (e.g. a finite duty-cycle, a
better modelling of the interaction between the cooling gas and radio jets -
see also \citealt{Fontanot_etal_2011}). Thus, there is room for residual star
formation at late times in BCGs, which would increase the metallicity of their
stellar population while decreasing their ages.  However, we have shown that
the stellar metallicities of the most massive galaxies would be too low with
respect to data even in the case AGN feedback is switched off. An
improved treatment of the satellite evolution that allows residual star
formation in satellite galaxies would help increasing the stellar metallicity
of massive galaxies. However, predictions from the \guo model (that uses such a
scheme) are still off the observational measurements.

Uncertainties in the stellar yields can affect model stellar
metallicities. However, the \dlb and \guo models assume already a relatively
large yield ($y=0.03$) while the \bow model assumes $y=0.02$ which is closer
but still somewhat larger than a `standard' value. In addition, one should
consider that increasing the yield would increase not only the total amount of
metals in the stars (and in the gas), but also the overall luminosities of
model galaxies (because of the strong dependence of cooling rates on
metallicity). A variable Initial Mass Function (IMF) could also affect the
total metal content of the most massive galaxies if these formed earlier than
their lower mass counterparts, from material with different physical
properties. For example, hydrodynamical simulations by
\citet{Smith_and_Sigurdsson_2007} show that above a critical metallicity of
about $10^{-3}\,{Z}_{\odot}$ clouds can fragment to form low-mass stars, while
for gas of lower metallicities stars form following a more top-heavy IMF. This
critical metallicity value, however, is well below that of observed
galaxies. Finally, our chemical enrichment scheme is rather crude, neglecting
the mass dependent lifetimes of stars, the influence of inefficient mixing and
metal loading. All these processes can affect the distribution of metals in
different baryonic components. To illustrate the importance of SN feedback, we
show in Figure~\ref{fig:check_noagn} predictions from a model with AGN feedback
but an alternative SN feedback (adopted in \citealt*{DeLucia_etal_2004}, based
on energy conservation arguments - purple cross hatched histograms). As
explained in \dlb, this model results in less efficient outflows and therefore
longer star formation histories and higher stellar metallicities. However, it
also over-predicts the overall luminosities of model galaxies. This example
highlights the sensitivity of metallicity on the details of SN feedback
models. Therefore, requiring a model to predict at the same time the correct
number densities, ages and metallicities of galaxies provides strong
constraints on the processes responsible for the gas and metal recycling, and
on the time scales involved.

In future work, we will investigate how results change when more detailed
chemical enrichment and recycling schemes are adopted.  Some fine-tuning might
be needed in order to avoid making the most massive galaxies too
young. Therefore, it appears also very important to obtain better stellar
population estimates for BCGs, and to understand how observational estimates
compare with the mass and/or luminosity weighted values from models.

\section*{Acknowledgements}
GDL acknowledges financial support from the European Research Council under the
European Community's Seventh Framework Programme (FP7/2007-2013)/ERC grant
agreement n. 202781. SB acknowledges support by the European Commission’s FP7
Marie Curie Initial Training Network CosmoComp (PITN-GA-2009-238356), the
PRIN-MIUR09 ``Tracing the growth of structures in the Universe'' and the PD51
INFN grant.  The Millennium Simulation databases used in this paper were
constructed as part of the activities of the German Astrophysical Virtual
Observatory. We acknowledge useful discussions with F. Fontanot, F. Matteucci,
P. Monaco, and S. Weinmann.

\bsp

\label{lastpage}

\bibliographystyle{mn2e}
\bibliography{met_bcgs}

\end{document}